\documentclass[prd,aps,showpacs,,nofootinbib,superscriptaddress,preprintnumbers,floatfix]{revtex4}
\usepackage[english]{babel}
\usepackage{pstricks}
\usepackage{graphicx}

\newcommand{\gev}{\,{\rm GeV}}

\newcommand{\AmS}{{\protect\the\textfont2
  A\kern-.1667em\lower.5ex\hbox{M}\kern-.125emS}}

\begin{document}
 \makeatletter
 \def\preprint#1{
 }
\preprint{\begin{tabular}{l}
      arXiv \\
      CPHT/S 
    \end{tabular}
 }
%
%
\hfill CPHT-RR002.0113

\hfill IRFU-13-01

\title{On timelike and spacelike  deeply virtual Compton scattering at next to leading order}
\author{H. Moutarde}
\affiliation{Irfu-SPhN, CEA, Saclay, France}
\author{B. Pire}
\affiliation{CPHT, {\'E}cole Polytechnique, CNRS, 91128 Palaiseau, France}
\author{F. Sabati\'e}
\affiliation{Irfu-SPhN, CEA, Saclay, France}
\author{L. Szymanowski}
\affiliation{National Center for Nuclear Research (NCBJ), Warsaw, Poland}
\author{J. Wagner}
\affiliation{National Center for Nuclear Research (NCBJ), Warsaw, Poland}

\begin{abstract}
We study timelike and spacelike  virtual Compton scattering  in the generalized
Bjorken scaling regime at next to leading order in the strong coupling constant, in the medium energy range which will be studied 
intensely at JLab12 and in the COMPASS-II experiment at CERN. 
We show that the Born amplitudes  get sizeable $O(\alpha_s)$ corrections and, even at moderate energies, the gluonic
contributions  are by no means negligible. We stress that the timelike and spacelike cases are complementary
and that their difference deserves much special attention.
\end{abstract}
%
\maketitle
\thispagestyle{empty}
\renewcommand{\thesection}{\arabic{section}}

\renewcommand{\thesubsection}{\arabic{subsection}}
\noindent
\section{Introduction}
Spacelike Deeply Virtual Compton Scattering (DVCS) \cite{historyofDVCS} 
\begin{equation}
\gamma^*(q_{in}) N(P) \to \gamma(q_{out}) N'(P'=P+\Delta) \,,~~~~~q_{in}^2 =-Q^2,~~~~~q_{out}^2 =0\,,
\end{equation}
has been the model reaction for studying QCD collinear factorization in exclusive processes in terms of Generalized Parton Distributions (GPDs) \cite{gpdrev1, gpdrev2}, which access correlated information about the light cone momentum fraction and transverse location \cite{Impact} of partons in hadrons. The specific kinematical regime where this factorization property is proven at leading twist is the generalized Bjorken regime of large energy and large $Q^2$ but finite and fixed momentum transfer $\Delta^2$. A number of experimental results at various energies \cite{expdvcs} have now established the relevance of this approach at accessible kinematical conditions. Detailed phenomenological studies \cite{fitting} are under way to quantify to which degree one may in a foreseeable future extract from experimental data the physical information encoded in GPDs. The DVCS process contributes to the leptoproduction of a real photon
\begin{equation}
l^\pm (k) N(P) \to l^\pm (k') \gamma(q_{out}) N'(P'=P+\Delta)\,\,.
\end{equation}
It interferes with the Bethe-Heitler process
\begin{equation}
l^\pm (k) \gamma^*(-\Delta) \to l^\pm (k') \gamma(q_{out}) \,\,,  
\end{equation}
where the hadronic interaction is entirely determined by the nucleon (spacelike) electromagnetic form factors $F_1(\Delta^2)$ and $F_2(\Delta^2)$.

Timelike Compton Scattering (TCS) \cite{BDP}
\begin{equation}
\gamma(q_{in}) N(P)\to \gamma^*(q_{out}) N'(P'=P+\Delta)\,,~~~~~q_{in}^2 =0,~~~~~q_{out}^2 =Q^2\,,
\end{equation}
which contributes to the photoproduction of a lepton pair
\begin{equation}
\gamma(q_{in}) N(P) \to l^- (k)  l^+ (k') N'(P'=P+\Delta)\,,~~~~~k +k' =q_{out}\,,
\end{equation}
and interferes with the Bethe-Heitler process
\begin{equation}
\gamma(q_{in})  \gamma^*(-\Delta) \to l^- (k)  l^+ (k') \,\,,  
\end{equation}
shares many features with DVCS and allows in principle to access the same GPDs.
The experimental situation \cite{exptcs} is not as encouraging as in the DVCS case but progress is expected in the next few years. The amplitudes of these two reactions are related at Born order by a simple complex conjugation but they significantly differ at next to leading order (NLO) in the strong coupling constant $\alpha_s$ \cite{MPSW}.

Complete NLO calculations \cite{JiO,Mankiewicz,Belitsky:1999sg,NLOphenoDVCS,Pire:2011st} are available for both the DVCS and TCS reactions and there is no indication that NLO corrections are negligible in the kinematics relevant for current or near future experiments.

In this paper, we explore the consequence of including NLO  gluon coefficient functions and NLO corrections to the quark coefficient functions entering the DVCS and TCS amplitudes, firstly in the calculation of spacelike and timelike Compton form factors with two models of GPDs. Then we proceed to the calculation of specific observables in the kinematical conditions which will soon be accessible in lepton nucleon collisions.

We focus on kinematics relevant to the next JLab and COMPASS measurements. The case of very large energy, and subsequently very small skewness, deserves its own study that will be addressed elsewhere. Ultraperipheral collisions at hadron colliders may already open the access to timelike Compton scattering in this domain \cite{PSW}. Also the proposed next generation of electron-ion colliders \cite{eic,eicwp} will access in a detailed way this interesting domain with small skewness.

In this paper we concentrate on the influence of the NLO corrections to the DVCS and TCS observables. Because of that, although required to ensure QED gauge invariance \cite{twist3}, we do not discuss here twist 3 effects. Neither do we implement the new results on target mass and finite $t$ corrections~\cite{Braun:2012bg} nor the recently proposed resummation  formula~\cite{APSW} which focuses on the regions near $x=\pm \xi$.
\section{Kinematics and amplitudes}
\subsection{Kinematics}

We introduce two light-like vectors $p$ and $n$ satisfying $p^2=0$, $n^2=0$ and $np=1$. We decompose the momenta in this light-cone basis as $k^\mu=\alpha n^\mu +\beta p^\mu +k_T^\mu$, with $k_T^2<0$, and we further note $k^+ = kn$. We also introduce the standard kinematical variables:
$\Delta=P'-P$, $t=\Delta^2$ and $W^2=(q_{in}+P)^2$.

We define kinematics separately for spacelike (sl) and timelike (tl) Compton scattering. We denote the (positive) skewness variable as $\xi$ in the DVCS case, and as $\eta$ in the TCS case.
In the DVCS case, where $-q_{in}^2=Q^2>0$ and $q_{out}^2=0$, we parametrize the momenta as follows
\begin{eqnarray}
\label{kinDVCS}
&& q_{in}^\mu = \frac{Q^2}{4\xi}n^\mu -2\xi p^\mu  \;, \;\;\;\;\;q_{out}^\mu = \alpha_{sl}n^\mu    -\frac{\Delta_T^2}{2\alpha_{sl}} p^\mu -\Delta_T^\mu\;,
\nonumber \\
&& P^\mu=(1+\xi)p^\mu+\frac{M^2}{2(1+\xi)}n^\mu\;,\;\;\;\; P'^\mu=\beta_{sl}p^\mu +\frac{M^2-\Delta^2_T}{2\beta_{sl}}n^\mu +\Delta_T^\mu \;,
\end{eqnarray}
where $M$ is the nucleon mass. The coefficients $\alpha_{sl}$, $\beta_{sl}$ in (\ref{kinDVCS}) satisfy the following system of equations
\begin{equation}
\label{eqsDVCS}
\alpha_{sl} = \frac{Q^2}{4\xi}+\frac{M^2}{2}\left(\frac{1}{1+\xi}-\frac{1}{\beta_{sl}} \right) +\frac{\Delta_T^2}{2\beta_{sl}}\;,\;\;\;\;\beta_{sl}= 1-\xi +\frac{\Delta_T^2}{2\alpha_{sl}}\;,
\end{equation}
which in the limit $M=0$ and $\Delta_T=0$ relevant for calculation of the coefficient function leads to the standard values $\alpha_{sl} = Q^2/4\xi$ and $\beta_{sl}= 1-\xi$.

In the TCS case, where $q_{in}^2=0$ and $q_{out}^2=Q^2>0$, we parametrize the momenta as
\begin{eqnarray}
\label{kinTCS}
&& q_{in}^\mu = \frac{Q^2}{4\eta}n^\mu  \;, \;\;\;\;\;q_{out}^\mu = \alpha_{tl}n^\mu  +\frac{Q^2-\Delta_T^2}{2\alpha_{tl}}p^\mu -\Delta_T^\mu\;,
\nonumber \\
&& P^\mu=(1+\eta)p^\mu+\frac{M^2}{2(1+\eta)}n^\mu\;,\;\;\;\; P'^\mu=\beta_{tl}p^\mu +\frac{M^2-\Delta^2_T}{2\beta_{tl}}n^\mu +\Delta_T^\mu \;.
\end{eqnarray}
The coefficients $\alpha_{tl}$, $\beta_{tl}$ in (\ref{kinTCS}) are solutions of the following system of equations
\begin{equation}
\label{eqsTCS}
\alpha_{tl} = \frac{Q^2}{4\eta}+\frac{M^2}{2}\left(\frac{1}{1+\eta}-\frac{1}{\beta_{tl}} \right) +\frac{\Delta_T^2}{2\beta_{tl}}\;,\;\;\;\;\beta_{tl}= 1+\eta-\frac{Q^2-\Delta_T^2}{2\alpha_{tl}}\;,
\end{equation}
which, again in the limit $M=0$ and $\Delta_T=0$ relevant for calculation of the coefficient function, take the standard values $\alpha_{tl} = Q^2/4\eta$ and $\beta_{tl}= 1-\eta$.

\subsection{The DVCS and TCS amplitudes}
After proper renormalization, the  full Compton scattering amplitude\footnote{We do not consider the photon helicity changing amplitude coming from the transversity gluon GPD \cite{transversity}.} reads in its factorized form (at factorization scale $\mu_F$)
\begin{eqnarray}
\mathcal{A}^{\mu\nu} &=& -g_T^{\mu\nu}\int_{-1}^1 dx 
\left[
\sum_q^{n_F} T^q(x) F^q(x)+T^g(x) F^g(x)
\right] \nonumber \\
&+& i\epsilon_T ^{\mu\nu}\int_{-1}^1 dx 
\left[
\sum_q^{n_F} \widetilde{T}^q(x) \widetilde{F}^q(x)+\widetilde{T}^g(x) \widetilde{F}^g(x)
\right] \,,
\label{eq:factorizedamplitude}
\end{eqnarray}
where we omitted the explicit skewness dependence. Renormalized coefficient functions are given by
\begin{eqnarray}
T^q(x)&=& \left[ C_{0}^q(x) +C_1^q(x) +\ln\left(\frac{Q^2}{\mu^2_F}\right) \cdot C_{coll}^q(x)\right] - ( x \to -x )  \,,\nonumber\\
T^g(x) &=& \left[ C_1^g(x) +\ln\left(\frac{Q^2}{\mu^2_F}\right) \cdot C_{coll}^g(x)\right] +( x \to -x )\,,\nonumber\\
\widetilde{T}^q(x)&=& \left[
\widetilde{C}_{0}^q(x) +\widetilde{C}_1^q(x) +\ln\left(\frac{Q^2}{\mu^2_F}\right) \cdot \widetilde{C}_{coll}^q (x)\right]
+( x \to -x )
\,,\nonumber\\
\widetilde{T}^g(x) &=&   \left[
\widetilde{C}_1^g(x) +\ln\left(\frac{Q^2}{\mu^2_F}\right) \cdot \widetilde{C}_{coll}^g(x)\right] - ( x \to -x )\,.
\label{eq:ceofficients}
\end{eqnarray} 
Results of the NLO calculations \cite{JiO,Mankiewicz,Belitsky:1999sg,NLOphenoDVCS,Pire:2011st} of the quark coefficient functions, based on the standard definitions of GPDs given in the Diehl's review \cite{gpdrev1}, read in the DVCS case
\begin{eqnarray}
C_0^q(x,\xi) &=& -e_q^2 \frac{1}{x+\xi-i\varepsilon} \,, \nonumber \\
C_1^q(x,\xi) &=& \frac{e_q^2\alpha_SC_F}{4\pi} \frac{1}{x+\xi-i\varepsilon}
\bigg[
9 -3\frac{x+\xi}{x-\xi}\log(\frac{x+\xi}{2\xi}-i\varepsilon)
-\log^2(\frac{x+\xi}{2\xi}-i\varepsilon )
\bigg]
\,, \nonumber\\
C_{coll}^q (x,\xi)&=&\frac{e_q^2\alpha_SC_F}{4\pi}
\frac{1}{x+\xi-i\varepsilon}
\bigg[
-3-2\log(\frac{x+\xi}{2\xi}-i\varepsilon) \bigg] 
\,, \nonumber\\
\widetilde{C}_0^q (x,\xi)&=& -e_q^2 \frac{1}{x+\xi-i\varepsilon} \,,\nonumber \\
\widetilde{C}_1^q (x,\xi)&=& \frac{e_q^2\alpha_SC_F}{4\pi} \frac{1}{x+\xi-i\varepsilon}
\bigg[
9 -\frac{x+\xi}{x-\xi}\log(\frac{x+\xi}{2\xi}-i\varepsilon)
-\log^2(\frac{x+\xi}{2\xi}-i\varepsilon )
\bigg]
\,, \nonumber\\
\widetilde{C}_{coll}^q (x,\xi)&=&\frac{e_q^2\alpha_SC_F}{4\pi}
\frac{1}{x+\xi-i\varepsilon}
\bigg[
-3-2\log(\frac{x+\xi}{2\xi}-i\varepsilon) \bigg] 
\,,
\label{eq:QDVCS}
\end{eqnarray}
where $C_F= (N_c^2-1) / (2N_c)$, $N_c$ being the number of colors, and $e_q$ is the quark electric charge in units of the proton charge. Using the same conventions, gluon coefficient functions read in the DVCS case
\begin{eqnarray}
C_1^g (x,\xi)&=& \frac{\Sigma e_q^2\alpha_ST_F}{4\pi}\frac{1}{(x+\xi-i\varepsilon)(x-\xi+i\varepsilon)}\times \nonumber \\
&&\bigg[
2\frac{x+3\xi}{x-\xi}\log\left(\frac{x+\xi}{2\xi}- i \varepsilon\right)
-\frac{x+\xi}{x-\xi}\log^2\left(\frac{x+\xi}{2\xi} - i \varepsilon\right)
\bigg] \,,\nonumber\\
C_{coll}^g (x,\xi)&=& \frac{\Sigma e_q^2\alpha_ST_F}{4\pi}\frac{2}{(x+\xi-i\varepsilon)(x-\xi+i\varepsilon)}\left[
-\frac{x+\xi}{x-\xi}\log\left(\frac{x+\xi}{2\xi} - i \varepsilon\right)
\right]\,, \nonumber \\
\widetilde{C}_1^g (x,\xi)&=& \frac{\Sigma e_q^2\alpha_ST_F}{4\pi}\frac{1}{(x+\xi-i\varepsilon)(x-\xi+i\varepsilon)}\times \nonumber \\
&&\bigg[
-2\frac{3x+\xi}{x-\xi}\log\left(\frac{x+\xi}{2\xi}- i \varepsilon\right)
+\frac{x+\xi}{x-\xi}\log^2\left(\frac{x+\xi}{2\xi} - i \varepsilon\right)
\bigg] \,,\nonumber\\
\widetilde{C}_{coll}^g (x,\xi)&=& \frac{\Sigma e_q^2\alpha_ST_F}{4\pi}\frac{2}{(x+\xi-i\varepsilon)(x-\xi+i\varepsilon)}\left[
\frac{x+\xi}{x-\xi}\log\left(\frac{x+\xi}{2\xi} - i \varepsilon\right)
\right]\,,
\label{eq:GDVCS}
\end{eqnarray}
where $T_F =\frac{1}{2}$. The results for the TCS case are simply \cite{MPSW}  related to these expressions 
\begin{eqnarray}
^{TCS}T(x,\eta) = \pm \left(^{DVCS}T(x,\xi=\eta) +  i \pi C_{coll}(x,\xi = \eta)\right)^* \,,
\label{eq:TCSvsDVCS}
\end{eqnarray}
where $+$~$(-)$ sign corresponds to vector (axial) case.

\section{Models for GPDs }
\label{GPD}
In our analysis we will use two GPD models based on Double Distributions (DDs) \cite{historyofDVCS,MR}. DDs allow to trivially achieve one of the strongest constraints on GPDs : the polynomiality of the Mellin moments of GPDs. They also automatically restore usual PDFs in the forward limit at $\xi,~t \to 0$. The GPDs are expressed as a two-dimensional integral over $\alpha$ and $\beta$ of the double distribution $f_i$
\begin{equation}
F_i (x,\xi,t) = \int_{-1}^1 d\beta \,  \int_{-1+|\beta |}^{1-|\beta |} d\alpha \, \delta (\beta+\xi\alpha - x) \,  f_i(\beta,\alpha,t) + D^F_i\left(\frac{x}{\xi},t\right)\, \Theta(\xi^2-x^2)\, ,
\label{Eq:GPD_DD}
\end{equation}
where $F = H, E, \widetilde{H}, \widetilde{E}$ and $i$ denotes the flavor ($val$ for valence quarks, $sea$ for sea quarks and $g$ for gluons). In our analysis we only take into account the contribution of $H$ and $\widetilde{H}$. Indeed, $E$ and $\widetilde{E}$ are mostly unknown, and recent phenomenological studies of Ref.~\cite{KMS} showed that most of existing DVCS observables are sensitive mostly to $H$ and $\widetilde{H}$.

The DD $f_i$ reads
\begin{equation}
f_i(\beta,\alpha,t) = g_i(\beta,t) \, h_i(\beta ) \, \frac{\Gamma (2n_i+2)}{2^{2n_i+1}\Gamma^2 (n_i+1)} \, \frac{[(1-|\beta |)^2-\alpha^2]^{n_i}}{(1-|\beta |)^{2n_i+1}}    \, ,
\label{Eq:DD}
\end{equation}
where $\Gamma$ is the gamma function, $n_i$ is set to 1 for valence quarks, 2 for sea quarks and gluons. $g_i(\beta,t)$ parametrizes the $t$-dependence of GPDs, the $h_i(\beta)$'s in case of GPDs $H$ and $\widetilde{H}$ denote their forward limit and are related to the usual polarized and unpolarized PDFs in the following way
\begin{eqnarray}
h_g( \beta ) & = & | \beta | \, g(| \beta |) \nonumber   \, , \\
h^q_{\rm sea}( \beta ) & = & q_{\rm sea} ( | \beta | ) \, {\rm sign}(\beta )   \, , \nonumber \\
h^q_{\rm val}( \beta ) & = & q_{\rm val}(  \beta  ) \, \Theta(\beta)  \, ,\nonumber\\
\tilde{h}_g( \beta ) & = & \beta  \, \Delta g(| \beta |) \nonumber   \, , \nonumber\\
\tilde{h}^q_{\rm sea}( \beta ) & = &  \Delta q_{\rm sea} ( | \beta | ) \, , \nonumber \\
\tilde{h}^q_{\rm val}( \beta ) & = & \Delta q_{\rm val}(  \beta  ) \, \Theta(\beta)  \, .
\end{eqnarray}
$D^F_i$ in Eq.~(\ref{Eq:GPD_DD}) denotes the Polyakov-Weiss D-term \cite{Polyakov:1999gs}. In our estimates we will use parametrizations obtained by a fit to the chiral soliton model \cite{KivPolVan}:
\begin{eqnarray}
D^H_q(\frac{x}{\xi},t) = - D^E_q(\frac{x}{\xi},t) &=& \frac{1}{3}D^q(\frac{x}{\xi})F_D(t) \, , \nonumber\\
D^H_g(\frac{x}{\xi},t) = - D^E_g(\frac{x}{\xi},t) &=& \xi D^g(\frac{x}{\xi})F_D(t) \, , 
\end{eqnarray}
where
\begin{eqnarray}
D^q(x,t) &=& - (1-x^2)\sum_{n=1}^{\infty(odd)} d_n^q~ C_n^{\frac{3}{2}}(x) \, , \nonumber \\
D^g(x,t) &=& - \frac{3}{2}(1-x^2)^2\sum_{n=1}^{\infty(odd)} d_n^g~ C_{n-1}^{\frac{5}{2}}(x) \, , 
\end{eqnarray}
where at $\mu =0.6 \gev$ matching to chiral soliton model gives: $d_1^q =4.0$, $d_3^q =1.2$, $d_5^q = 0.4$, and we make an assumption that $d^g_n$ at input scale vanishes. In the QCD evolution of $d^q_n, d^g_n$ we switch from 3 to 4 flavours at $\mu = 1.5 \gev$ and $\Lambda_3 =0.232 \gev$, $\Lambda_4 =0.200 \gev$.


\subsection{The Goloskokov-Kroll model for the GPDs }
\label{secGK}
As described in details in Refs.~\cite{GK1,GK2,GK3,KMS}, the GPDs of the so-called Goloskokov-Kroll (or GK) model is constructed using CTEQ6m PDFs \cite{cteq6m}. The low-$x$ behavior of PDFs is well reproduced by power-law, with the power assumed to be generated by Regge poles. In this GPD model, the Regge behavior with linear trajectories of the DD encoded in the function $g_i(\beta,t)$ (see Eq.~(\ref{Eq:DD})) is assumed
\begin{equation}
g_i(\beta,t) = {\rm e}^{b_i t} \, | \beta |^{-\alpha_i' t}  \, .
\label{Eq:KG_Regge}
\end{equation}
%
For the unpolarized GPD $H(x,\xi,t)$ the values of the Regge trajectory slopes and residues $\alpha_i', b_i$  as well as the expansions of the CTEQ6m PDFs \cite{cteq6m} used for $h_i$ may be found in Refs.~\cite{GK1,GK2,GK3}. 
Finally, this model uses simple relations to parameterize the quark sea
\begin{eqnarray}
H^u_{\rm sea} &=& H^d_{\rm sea} =  \kappa_s H^s_{\rm sea} \, , \nonumber \\
{\rm with} \;\;\;\;\;\; \kappa_s&=&1+0.68/(1+0.52 \ln Q^2/Q_0^2 ) \, ,
\end{eqnarray}

\noindent with the initial scale of the CTEQ6m PDFs $Q_0^2=4$~GeV$^2$.

Similarly to the GPD $H$, polarized GPD $\widetilde H$ is constructed using the Bl\"umlein - B\"ottcher (BB) polarized PDF parametrization \cite{BB} to fix the forward limit. Meson electroproduction data from HERA and HERMES have been considered to fix parameters for this GPD in the GK model, $\widetilde H$ for valence quarks and gluons have been parametrized, however we have neglected $\widetilde H$ for sea quarks. The values of the Regge trajectory slopes and residues  as well as the expansions of the BB PDFs used in the GK model may be found in Refs.~\cite{GK1,GK5} . 

\subsection{The MSTW08 based model with factorized $t$ - dependence} 

For the second model we use double distribution with MSTW08 PDFs \cite{MSTW08}.  In that case we take simple factorizing ansatz for $t$ - dependence
\begin{eqnarray}
g_u(\beta , t) &=& \frac{1}{2}F_1^u(t)\, , \\
g_d(\beta , t) &=& F_1^d(t)\, , \\
g_s(\beta , t) &=& g_g(\beta , t) = F_D(t)\, ,
\end{eqnarray}
where
\begin{eqnarray}
F_1^u(t) &=& 2 F_1^p(t) +  F_1^n(t)\, , \\
F_1^d(t) &=&   F_1^p(t) +2 F_1^n(t)\, , \\
F_D(t) 	 &=&  (1 -t/M_V^2)^{-2}\,,
\end{eqnarray}
with $M_V =0.84 \gev$, $F_1^p$ and $F_1^n$ are electromagnetic Dirac spacelike form factors of the proton and neutron. We use that model to construct only $\mathcal{H}$.
\section{Compton Form Factors}


Let us now present the results for spacelike and timelike Compton Form Factors (CFF) at NLO,  $\mathcal{H}$ and $\widetilde \mathcal{H}$, defined in the DVCS case as
\begin{eqnarray}
\mathcal{H}(\xi,t) &=& + \int_{-1}^1 dx \,
\left(\sum_q T^q(x,\xi)H^q(x,\xi,t)
 + T^g(x,\xi)H^g(x,\xi,t)\right) \nonumber \\
\widetilde \mathcal{H}(\xi,t) &=& - \int_{-1}^1 dx \,
\left(\sum_q \widetilde T^q(x,\xi)\widetilde H^q(x,\xi,t) 
+\widetilde T^g(x,\xi)\widetilde H^g(x,\xi,t)\right).
\label{eq:CFF}
\end{eqnarray}

These CFFs are the GPD dependent quantities which enter the amplitudes. For DVCS they are defined through relations such as \cite{BDP} 
\begin{equation}
\mathcal{A}^{\mu\nu}(\xi,t) = - e^2 \frac{1}{(P+P')^+}\, \bar{u}(P^{\prime}) 
\left[\,
   g_T^{\mu\nu} \, \Big(
      {\mathcal{H}(\xi,t)} \, \gamma^+ +
      {\mathcal{E}(\xi,t)} \, \frac{i \sigma^{+\rho}\Delta_{\rho}}{2 M}
   \Big)
   +i\epsilon_T^{\mu\nu}\, \Big(
    {\widetilde{\mathcal{H}}(\xi,t)} \, \gamma^+\gamma_5 +
      {\widetilde{\mathcal{E}}(\xi,t)} \, \frac{\Delta^{+}\gamma_5}{2 M}
    \Big)
\,\right] u(P) \, ,
\label{eq:amplCFF}
\end{equation}
Similar relation holds for TCS with $\xi$ replaced by $\eta$.

We now present our results for the spacelike and timelike virtual Compton scattering, for the $\xi$ and $\eta$ values which include kinematical regimes of the JLab and Compass experiments. As we present our results for $Q^2 = \mu_F^2 = \mu_R^2 = 4 $ GeV$^2$, throughout the whole paper we use the value of $\alpha_S = 0.3$~.

\subsection{Spacelike Compton Form Factors}

\begin{figure}[p]
\begin{center}
  \includegraphics[width= 0.85\textwidth]{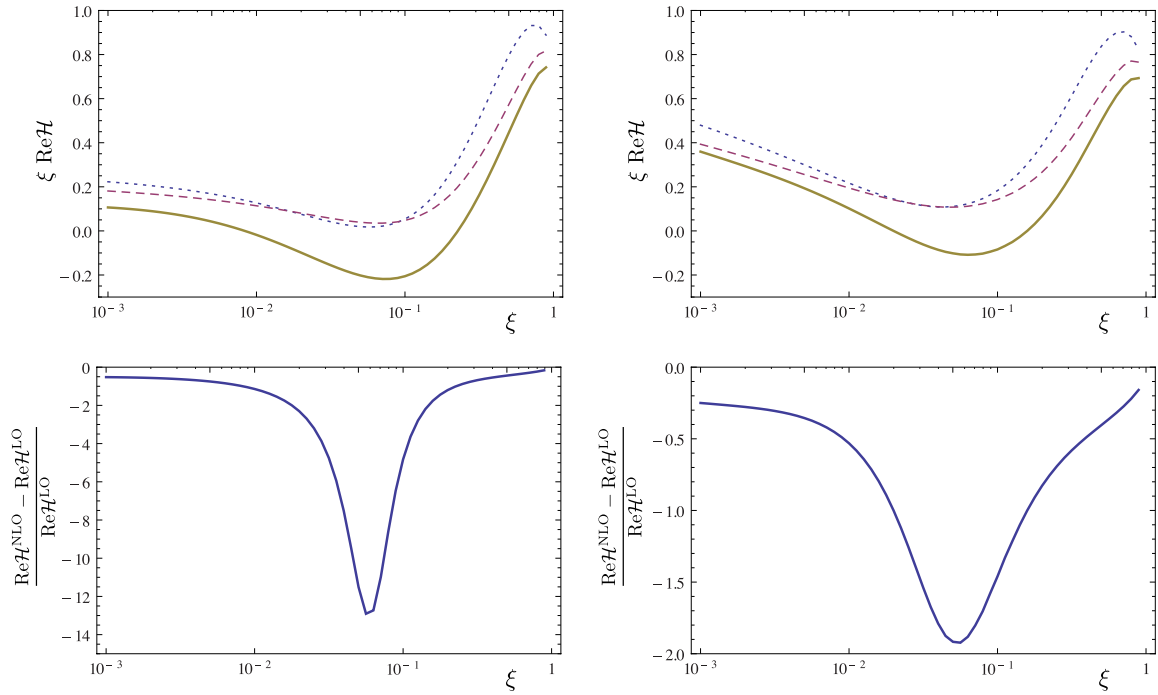} 
\caption{The real part of the {\it spacelike} Compton Form Factor $\mathcal{H}(\xi)$ multiplied by $\xi$, as a function of $\xi$ in the double distribution model based on Kroll-Goloskokov (upper left) and MSTW08 (upper right) parametrizations, for $
\mu_F^2=Q^2=4 \gev^2$ and $t= -0.1 \gev^2$, at the Born order (dotted line), including the NLO quark corrections (dashed line) and including both quark and gluon NLO corrections (solid line). Below the ratios of the NLO correction to LO result in the corresponding models.}
\label{fig:DVCSRe2x2}
\end{center}
\end{figure}
\begin{figure}[p]
\begin{center}
  \includegraphics[width= 0.85\textwidth]{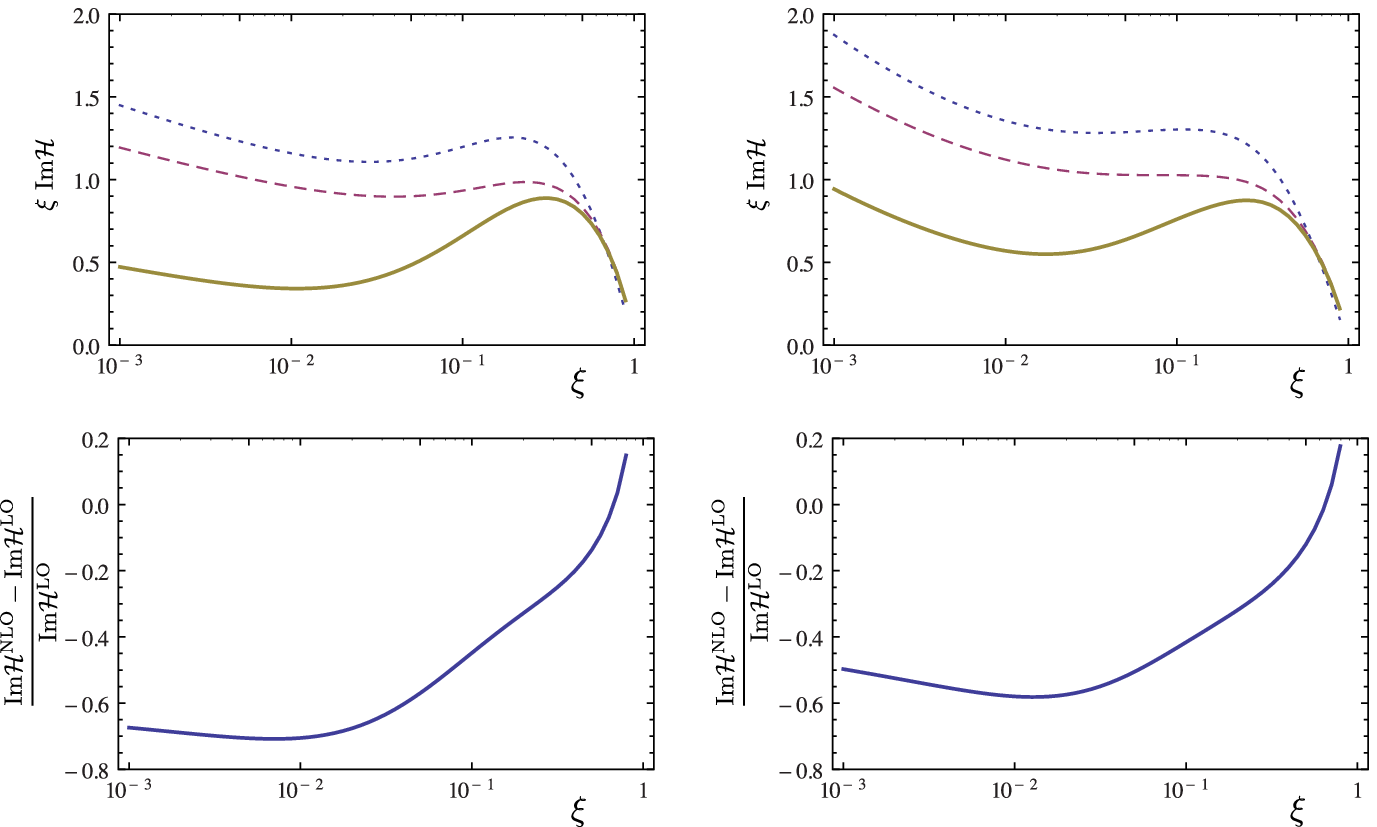} 
\caption{The imaginary part of the {\it spacelike} Compton Form Factor $\mathcal{H}(\xi)$ multiplied by $\xi$, as a function of $\xi$ in the double distribution model based on Kroll-Goloskokov (upper left) and MSTW08 (upper right) parametrizations, for $
\mu_F^2=Q^2=4 \gev^2$ and $t= -0.1 \gev^2$, at the Born order (dotted line), including the NLO quark corrections (dashed line) and including both quark and gluon NLO corrections (solid line). Below the ratios of the NLO correction to LO result in the corresponding models.}
\label{fig:DVCSIm2x2}
\end{center}
\end{figure}

Let us first discuss the importance of including NLO effects in CFFs related to DVCS observables. We show on Fig.~\ref{fig:DVCSRe2x2} (resp. Fig.~\ref{fig:DVCSIm2x2}) the results of our calculations for the real (resp. imaginary) parts of the CFF  $\mathcal{H}(\xi)$ (multiplied by $\xi$ for a better legibility of the figure, since the CFFs are roughly proportional to $1/\xi$), with the two GPD models described in Sect.~\ref{GPD}. The dotted curves are the LO results and the solid lines show the  results including all NLO effects. Although the results are naturally dependent of the choice of model GPDs, the main conclusions are quite universal. One can first observe that NLO corrections are by no means small, as exemplified by the ratio of these NLO corrections to the LO result shown in the lower part of Figs. \ref{fig:DVCSRe2x2} and  \ref{fig:DVCSIm2x2}. One can also observe that the NLO corrections tend to diminish the real part of the CFF, and even change its sign for $\xi \gtrsim 0.01$). NLO corrections also decrease the imaginary part of the CFF. These are not  new results \cite{Mankiewicz,Belitsky:1999sg,NLOphenoDVCS}. To quantify the main source of the NLO contribution, we show on the same plots with dashed lines the real and the imaginary parts of the CFF including quark NLO effects but not the gluon effects. Gluon effects are the most important part of the NLO correction, even at  quite large values of $\xi$ (up to around $\xi \approx 0.3$) and they  contribute as a very significant part to the full  CFF  $\mathcal{H}(\xi)$ including NLO effects. Since the CFF  $\mathcal{H}(\xi)$ dominates the DVCS amplitude, this means that extracting quark GPDs from a leading order analysis of DVCS data is, to say the least, questionable. More positively, this result indicates that DVCS experiments even in the low energy regime of JLab12  provide us with a nice way to measure gluon GPDs. This fact has, to our knowledge, never been clearly spelled out before.

\begin{figure}[ht]
\begin{center}
  \includegraphics[width= 0.85\textwidth]{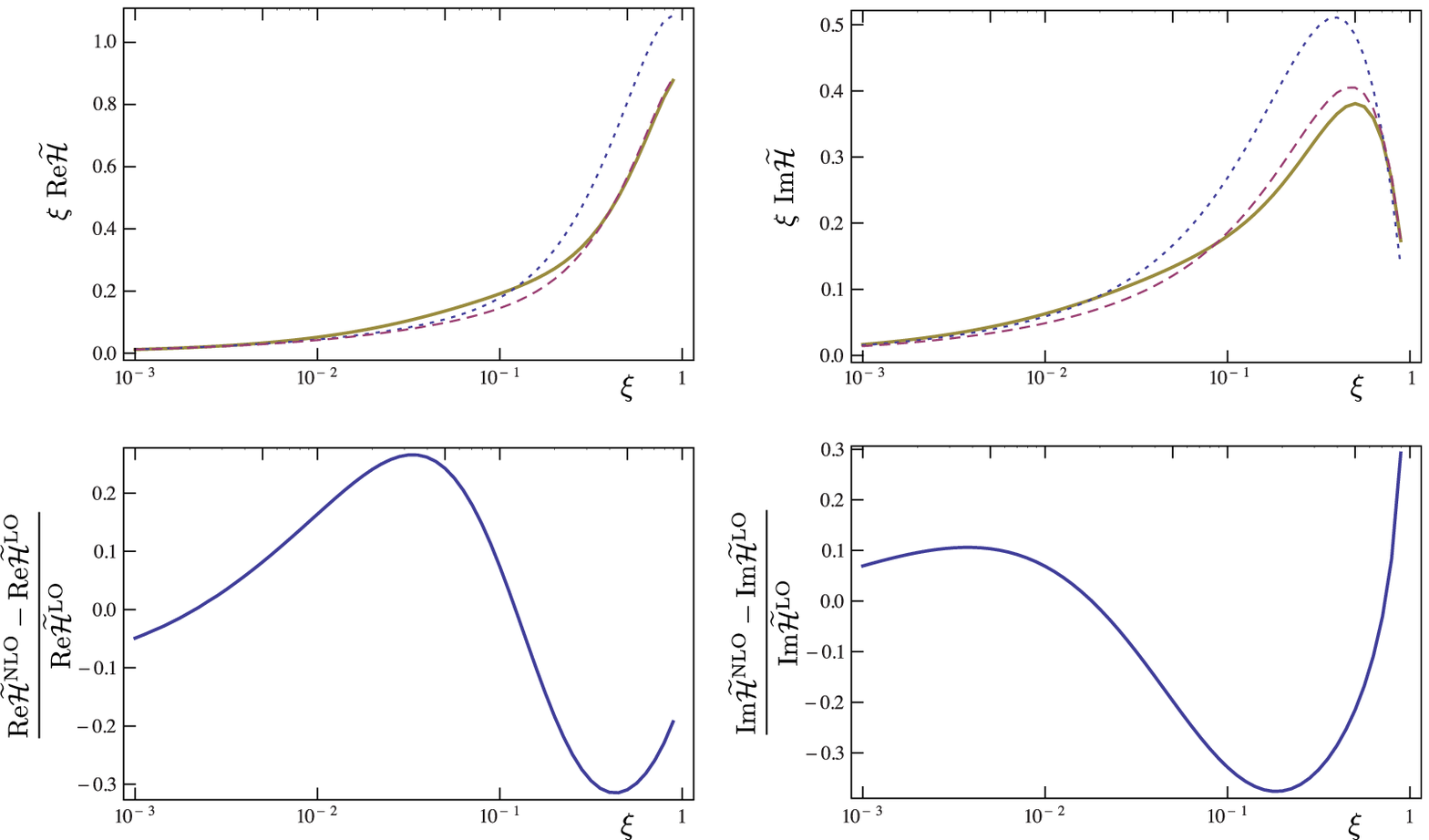}
\caption{The  real (left) and imaginary(right) parts of the {\it spacelike} Compton Form Factor $\widetilde \mathcal{H}$ multiplied by $\xi$  as a function of $\xi$ in the double distribution model based on KG  parametrizations, for $\mu_F^2=Q^2=4$~GeV$^2$ and $t= -0.1$~GeV$^2$, at the Born order (dotted line), including the NLO quark corrections (dashed line) and including both quark and gluon NLO corrections (solid line). The plots in the lower part show the corresponding  ratios of the NLO correction to the LO results.}
\label{fig:DVCStH}
\end{center}
\end{figure}

For completeness, we show on Fig.~\ref{fig:DVCStH} the real and imaginary parts of the CFF $\xi \widetilde \mathcal{H} (\xi)$ only for KG model. As for the case of $ \mathcal{H} (\xi)$, although the effects are less dramatic here, the NLO corrections are by no means small. 


Let us now comment on the D-term contribution to the CFFs. Since the GPD originating from a D-term is a function of the ratio $x/\xi$ with a support in the ERBL region $-\xi < x < \xi$, it results in a constant real contribution to CFFs in the spacelike case. For $Q^2= \mu_F^2 = 4$ GeV$^2$ and $t = -0.1$ GeV$^2$ , the values of different contributions to $\mathcal{H_D}$ are shown in the second column of the Tab.~\ref{tableDTCS}. One can see that this D-term contribution is very important for $\xi \sim 0.1$, since it modifies the  the full NLO result by  about $50\%$.  It is much less important at lower values of $\xi$ (around $3\%$ for $\xi \sim 0.01$ and only a few per mil  at $\xi\sim 0.001$).

\subsection{Timelike Compton Form Factors}

\begin{figure}[p]
\begin{center}
  \includegraphics[width= 0.85\textwidth]{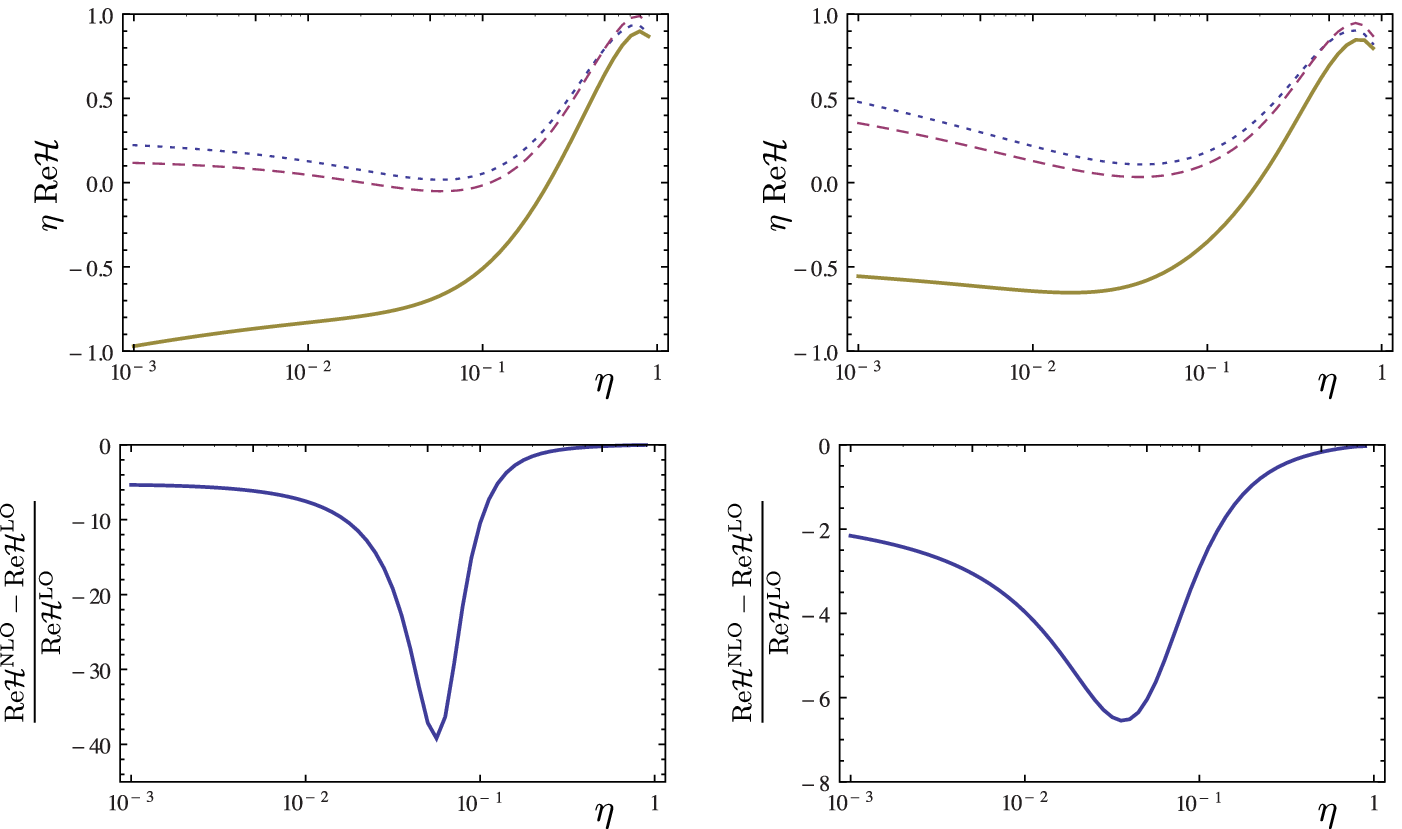} 
\caption{The real part of the {\it timelike} Compton Form Factor $\mathcal{H}$ multiplied by $\eta$, as a function of $\eta$ in the double distribution model based on Kroll-Goloskokov (upper left) and MSTW08 (upper right) parametrizations, for $\mu_F^2=Q^2=4$~GeV$^2$ and $t=-0.1$~GeV$^2$. Below the ratios of the NLO correction to LO result of the corresponding models.}
\label{fig:TCSRe2x2}
\end{center}
\end{figure}

\begin{figure}[p]
\begin{center}
  \includegraphics[width= 0.85\textwidth]{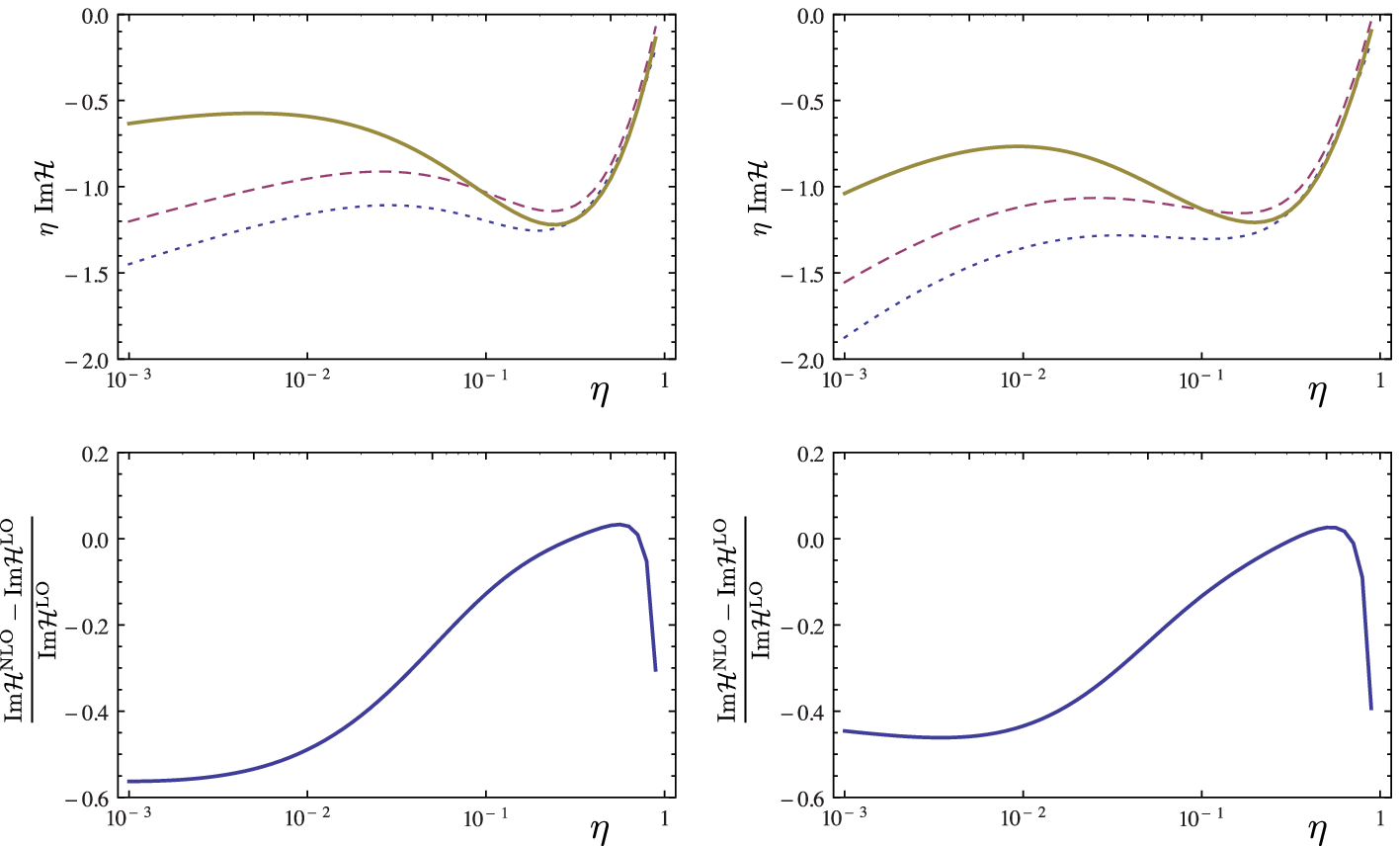} 
\caption{The imaginary part of the {\it timelike} Compton Form Factor $\mathcal{H}$ multiplied by $\eta$, as a function of $\eta$ in the double distribution model based on Kroll-Goloskokov (upper left) and MSTW08 (upper right) parametrizations, for $\mu_F^2=Q^2=4$~GeV$^2$ and $t= -0.1$~GeV$^2$. Below the ratios of the NLO correction to LO result of the corresponding models.}
\label{fig:TCSIm2x2}
\end{center}
\end{figure}

\begin{figure}[ht]
\begin{center}
  \includegraphics[width=7cm]{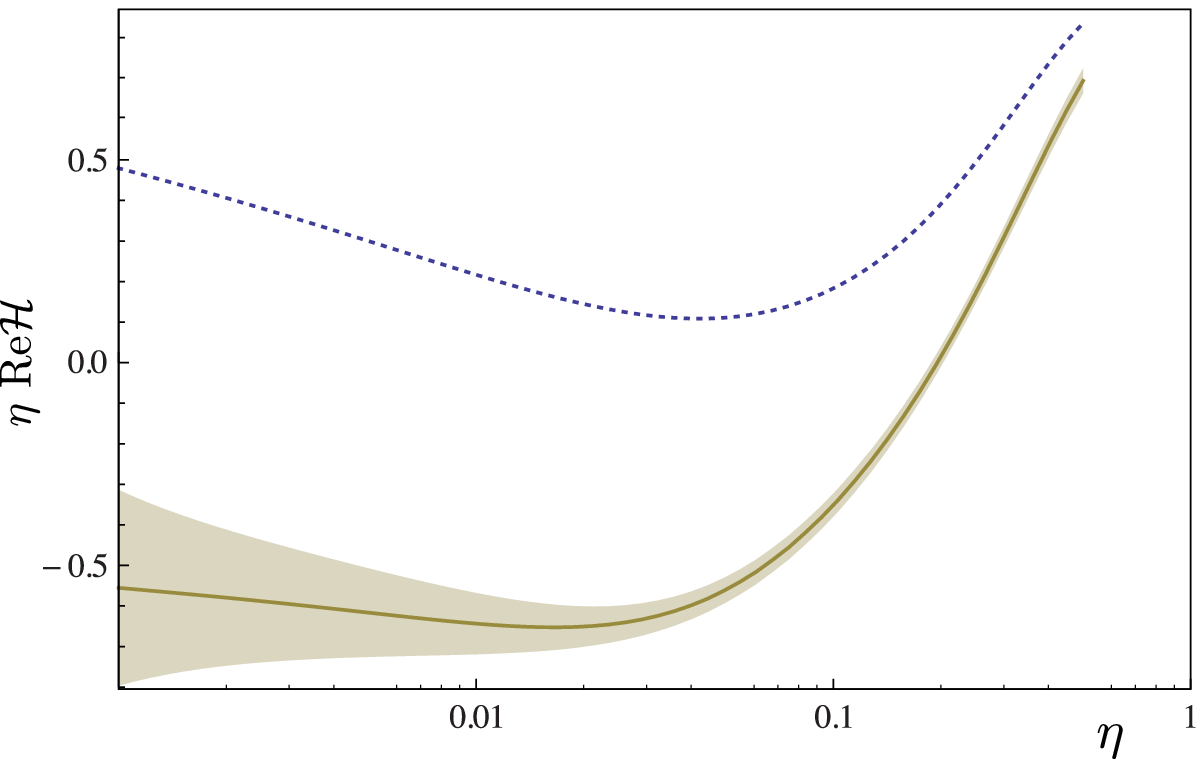}  ~~~~ \includegraphics[width= 7cm]{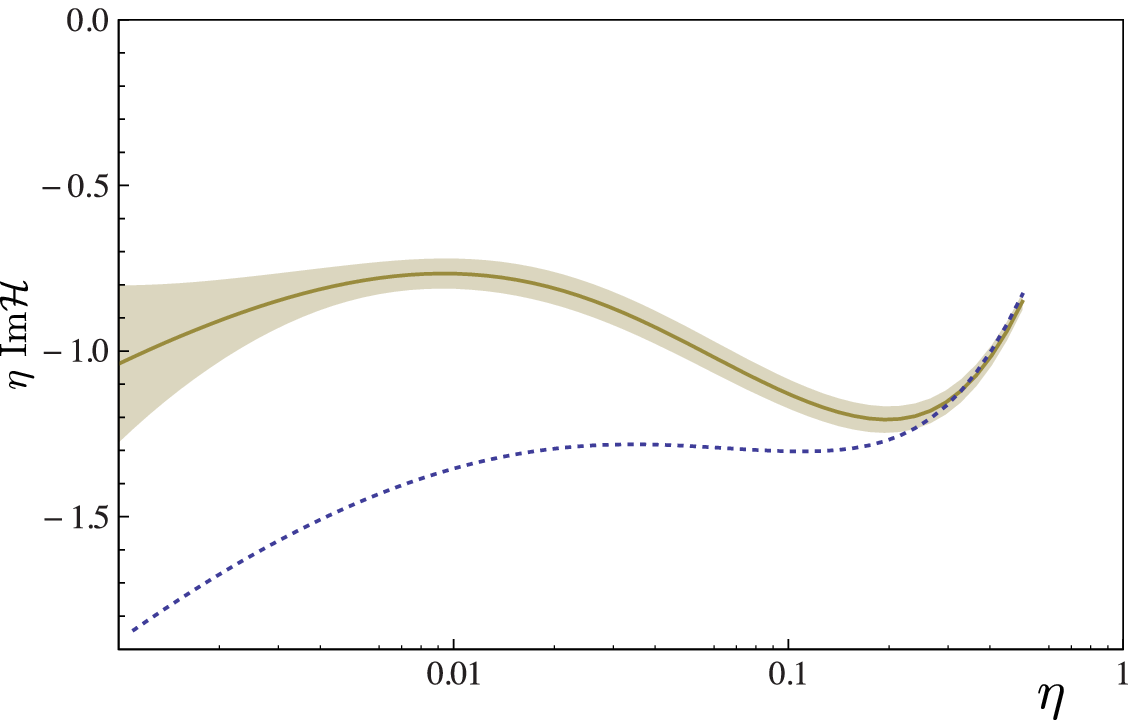} 
\caption{The real (left) and imaginary(right) parts of the TCS Compton Form Factor $\mathcal{H}$ multiplied by $\eta$, as a function of $\eta$ in the double distribution model based on MSTW08 parametrization, for $\mu_F^2=Q^2=4$~GeV$^2$ and $t= -0.1$~GeV$^2$. The dotted line shows the LO result and shaded bands around solid lines show the effect of a one sigma uncertainty of the input MSTW08 fit to the full NLO result.}
\label{fig:TCSerror}
\end{center}
\end{figure}

Let us now discuss the corresponding results for the timelike CFFs.   
On  Fig.~\ref{fig:TCSRe2x2} and Fig.~\ref{fig:TCSIm2x2}, we show the real and imaginary parts of the CFF $\mathcal H$ for the KG and the MSTW08 models of GPDs described in Sect.~\ref{GPD}, for the invariant mass of the lepton pair $Q'^2=4 \textrm{~GeV}^2$, $t=-.1\gev^2$ and factorization scale $\mu_F=Q$. For the imaginary part the correction does not exceed $40\%$. In the real part, the correction is of the order of few hundred percent. We observe that the main part of that large correction comes from the contribution of gluonic GPDs. To quantify the sensitivity of this statement on the uncertainties of the input PDF parametrizations we replot on the Fig. \ref{fig:TCSerror} the upper right panel of Fig. \ref{fig:TCSRe2x2} and the upper right panel of Fig. \ref{fig:TCSIm2x2}, now with the shaded bands showing the effect of a one sigma uncertainty of the input MSTW08 fit on the full NLO result, compared to the LO.
\begin{figure}[ht]
\begin{center}
  \includegraphics[width= 0.85\textwidth]{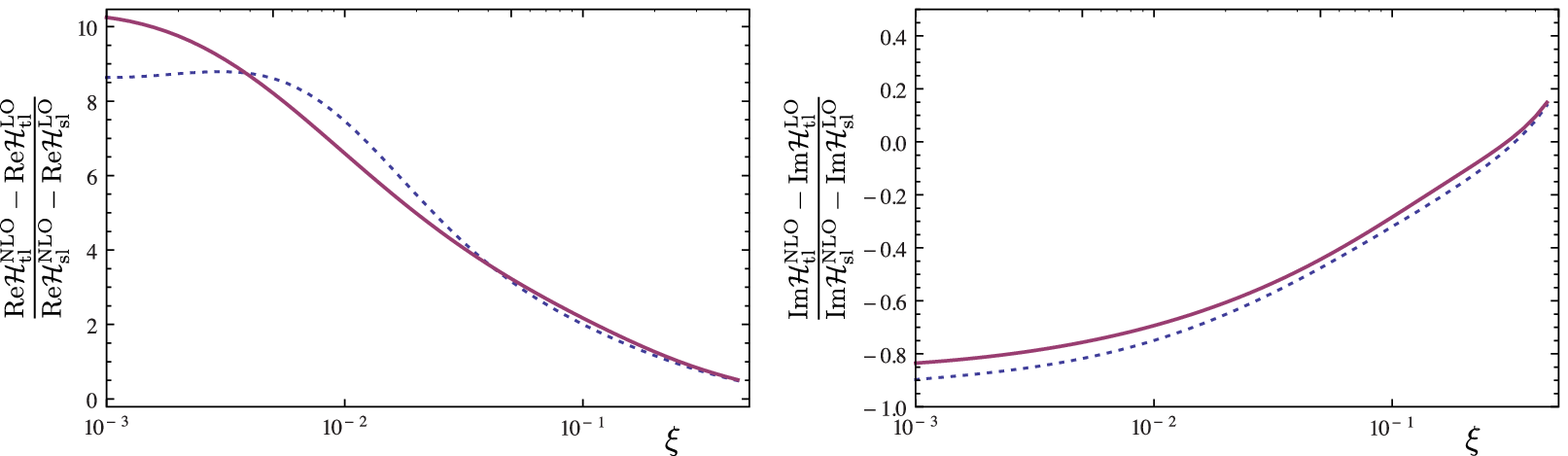}
\caption{The ratio of the timelike to spacelike NLO corrections in the real (left) and imaginary (right) part of the Compton Form Factor $\mathcal{H}$, as a function of $\xi$ in the double distribution model based on Kroll-Goloskokov (dashed) and MSTW08 (solid) parametrizations, for $\mu_F^2=Q^2=4$~GeV$^2$ and $t= -0.1$~GeV$^2$. For comparison timelike CFFs where calculated at $\eta = \xi$. }
\label{fig:TCStoDVCS}
\end{center}
\end{figure}

The D-term contribution to the CFF is a $\eta$-independent quantity and it has both a real and an imaginary parts at NLO in the TCS case. 
We show in Tab.~\ref{tableDTCS} the values of this D-term contribution in the LO and NLO cases. Its relative effect on the imaginary part of the CFF decreases significantly when $\eta$ decreases, from 10 to 1 and 0.1\%  when $\eta$ decreases from 0.1 to 0.01 and to 0.001.
\begin{table*}
\begin{center}
\begin{tabular}{|c||c|c|}
\hline
 & Re$\mathcal{H_D}$ &  Im$\mathcal{H_D}$ \\ \hline \hline
LO &  -2.59 & 0 \\ \hline
NLO quark contribution & -0.16 & -0.85 \\ \hline
NLO gluon contribution & 0.18 & 0.16 \\ \hline
Full NLO & -2.57 & -0.69 \\ 
\hline
\end{tabular}
\caption{\small Different contributions to the D-term. The values of the real part coincides for spacelike and timelike CFF $\mathcal H$, while the imaginary part is non-vanishing only for the timelike case.}
\label{tableDTCS}
\end{center}
\end{table*}


We then compare TCS and DVCS by plotting the ratio of NLO corrections on Fig.~\ref{fig:TCStoDVCS}. There is a striking difference in the magnitude of the corrections to the real part of CFFs, mostly insensitive to the choice of GPD parametrizations. As discussed in Ref.~\cite{MPSW}, this is a consequence of Eq.~(\ref{eq:TCSvsDVCS}) which by adding  a phase to the dominant imaginary part of the spacelike CFF at small skewness, gives rise to a sizeable real part of the corresponding CFF in the timelike case.
Such large corrections to the real part of CFFs will have a significant influence on observables which depend on the interference of the TCS process with the Bethe-Heitler amplitude, {\sl i.e.} connected to the azimuthal angular distribution of the leptons. We shall discuss this in the next section.

\begin{figure}[ht]
\begin{center}
  \includegraphics[width= 0.85\textwidth]{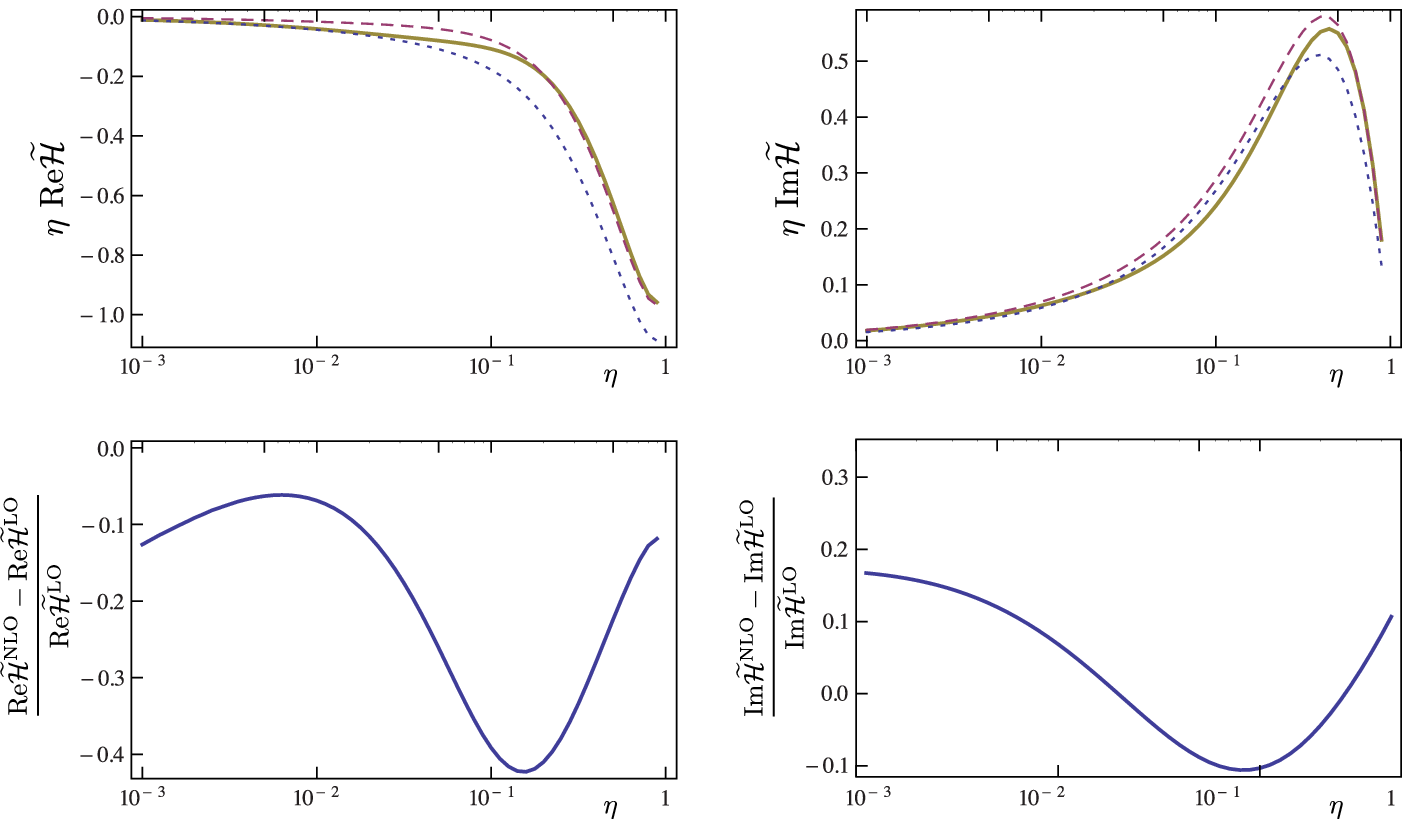}
\caption{The  real (left) and imaginary(right) parts of the {\it timelike} Compton Form Factor $\widetilde \mathcal{H}$ multiplied by $\eta$  as a function of $\eta$ in the double distribution model based on KG  parametrizations, for $\mu_F^2=Q^2=4$~GeV$^2$ and $t= -0.1$~GeV$^2$, at the Born order (dotted line), including the NLO quark corrections (dashed line) and including both quark and gluon NLO corrections (solid line). The plots in the lower part show the corresponding  ratios of the NLO correction to the LO results.}
\label{fig:TCStH}
\end{center}
\end{figure}

For completeness, we show on Fig.~\ref{fig:TCStH} the real and imaginary parts of $\eta \widetilde \mathcal{H} (\eta)$. The NLO corrections are here smaller than 20\% in the whole domain.

\section{Cross sections and asymmetries}
\subsection{Deeply virtual Compton scattering}

\begin{figure}[p]
\begin{center}
  \includegraphics[width=12.5cm]{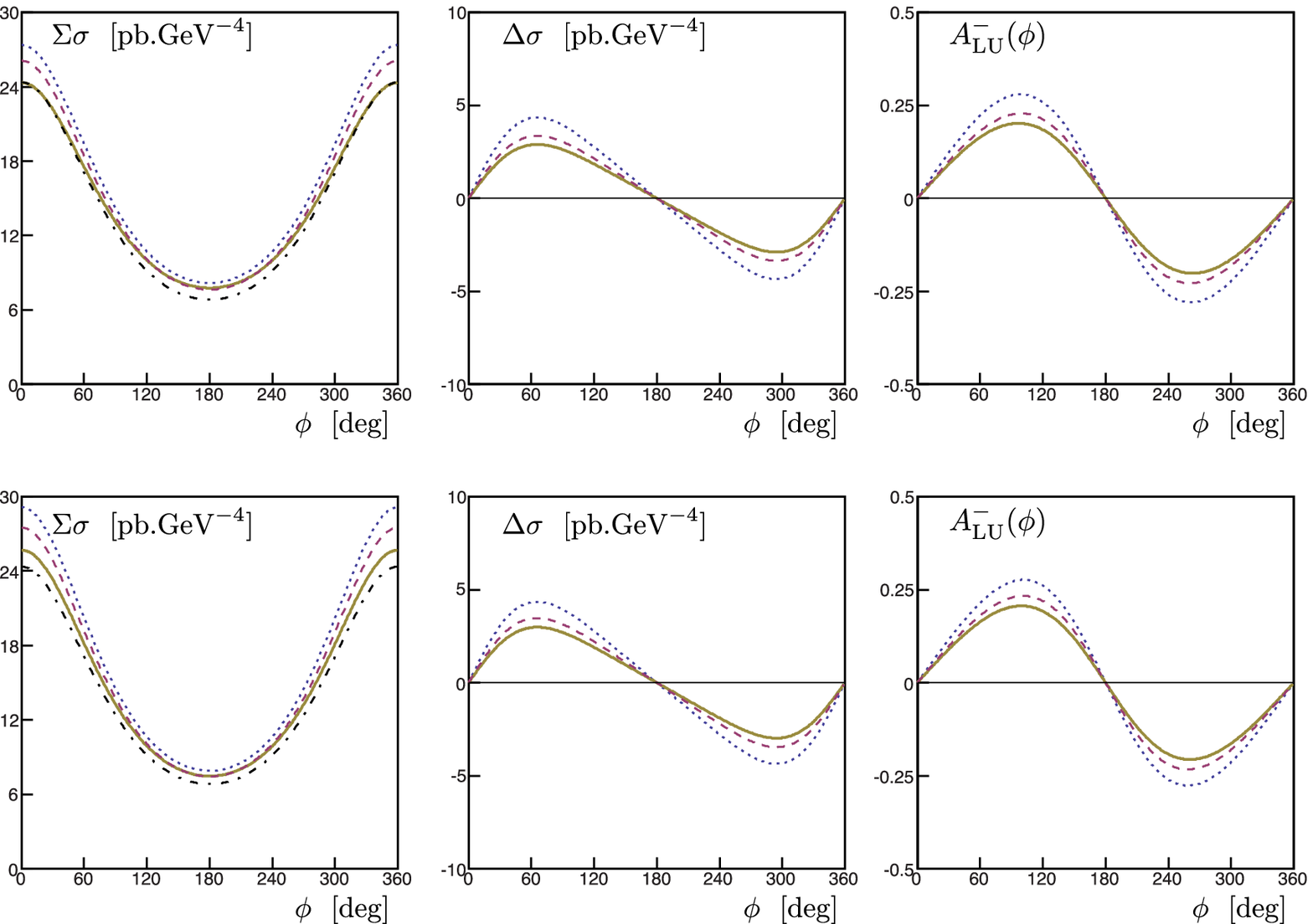}  
\caption{From left to right, the total DVCS cross section in pb/GeV$^4$, the difference of cross sections for opposite lepton helicities in pb/GeV$^4$, the corresponding asymmetry, all as a function of the usual $\phi $ angle (in the Trento convention \cite{Bacchetta:2004jz})  for $E_e=11 \gev, \mu_F^2=Q^2=4$~GeV$^2$ and $t= -0.2$~GeV$^2$. On the first line, the GPD $H(x,\xi,t)$ is parametrized by the GK model, on the second line  $H(x,\xi,t)$ is parametrized in the double distribution model based on the MSTW08 parametrization. The contributions from other GPDs are not included. In all plots, the LO result is shown as the dotted line, the full NLO result by the solid line and the NLO result without the gluonic contribution as the dashed line. The Bethe-Heitler contribution appears as the dash-dotted line in the cross section plots (left part).}
\label{fig:c1}
\end{center}
\end{figure}

Let us first briefly review the  effects of including NLO corrections on the DVCS observables. On Fig.~\ref{fig:c1} we show the total DVCS cross section, the difference of cross sections for opposite lepton helicities and  the beam spin asymmetry $A_{LU}^-$, defined by Eq.~(48) of Ref.~\cite{KMS}. We choose a set of values of kinematic variables representative for JLab12, namely $Q^2 = 4$~GeV$^2$, $E_{e}=11$~GeV and $t= -0.2$~GeV$^2$. The observables are shown  as a function of the azimuthal angle $\phi$ (in the Trento convention). The Born order result is shown as the dotted line, the full NLO result by the solid line and the NLO result without the gluonic contribution as the dashed line. We see that the effects of the NLO corrections are quite large in both GPD models. Although the value of $\xi$ is quite large, we see that the gluon contribution ({\it i.e.} the difference between the dashed and the solid curve) is by no means negligible.

\begin{figure}[p]
\begin{center}
  \includegraphics[width=12.5cm]{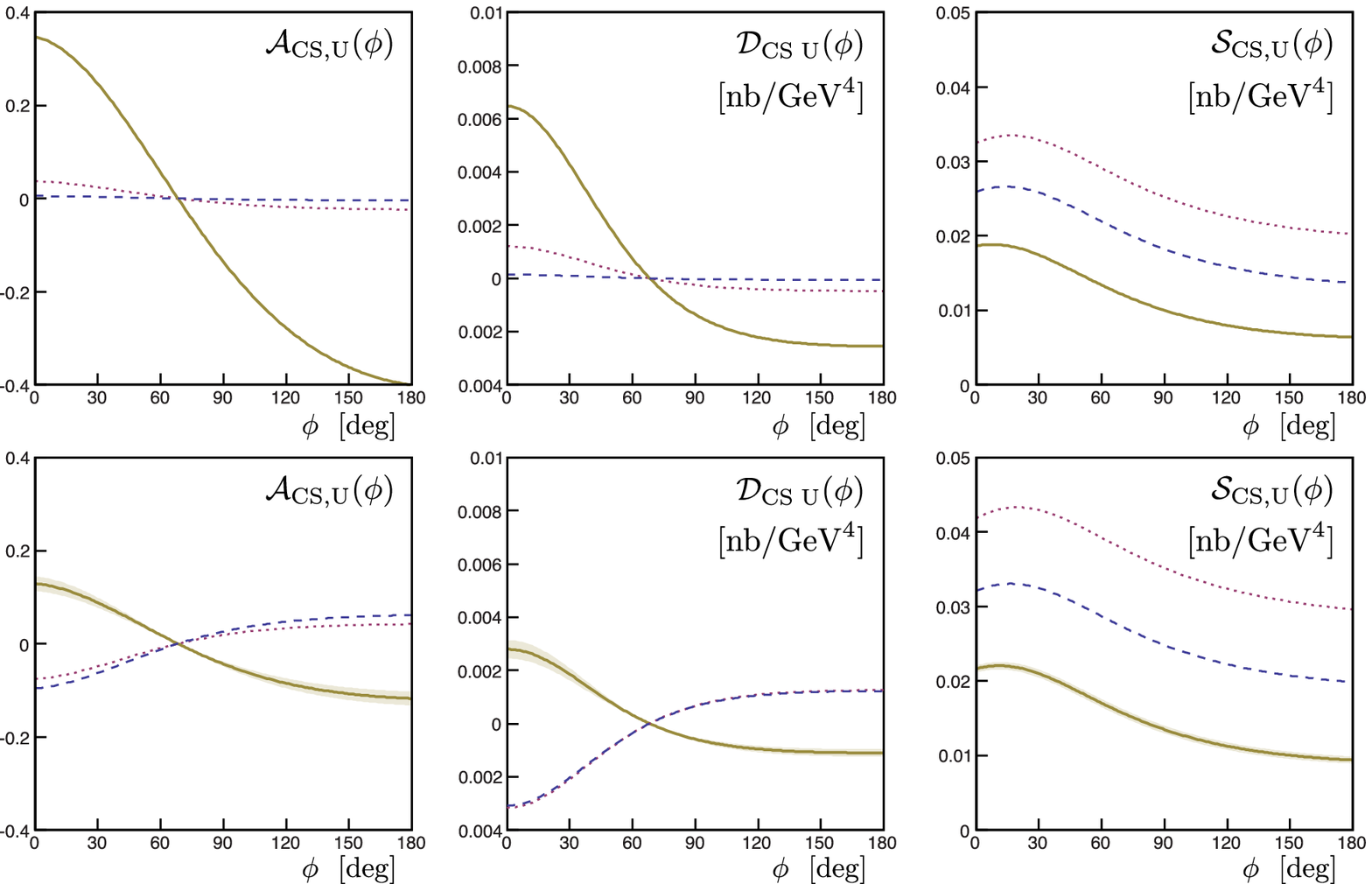}  
\caption{The DVCS observables for the COMPASS experiment, from left to right, mixed charge-spin asymmetry, mixed charge-spin difference and mixed charge-spin sum (in nb/GeV$^4$). The kinematical point is chosen as $ \xi=0.05, Q^2=4$~GeV$^2$, $t=-0.2$~GeV$^2$.  On the first line, the GPD $H(x,\xi,t)$ is parametrized by the GK model, on the second line  $H(x,\xi,t)$ is parametrized in the double distribution model based on the MSTW08 parametrization. The contributions from other GPDs are not included. In all plots, the LO result is shown as the dotted line, the full NLO result by the solid line and the NLO result without the gluonic contribution as the dashed line.}
\label{fig:compass}
\end{center}
\end{figure}

On Fig.~\ref{fig:compass} we show the DVCS observables relevant to the COMPASS experiment at CERN, namely (from left to right) the mixed charge-spin asymmetry, the mixed charge-spin difference and the mixed charge-spin sum defined in Eq. (59) of Ref.~\cite{KMS}, at the kinematics $\xi=0.05$, $Q^2=4$~GeV$^2$ and $-t=0.2$~GeV$^2$.
The upper part of  Fig.~\ref{fig:compass}  uses the GK parametrization and the lower part the MSTW08 parametrizations with $1\sigma$  errors. We display here only the contribution from the GPD $H$ . The lower value of the skewness $\xi=0.05$  allows to test a complementary regime with respect to JLab measurements.  Note the dramatic difference in the real part once gluon GPDs are in, inducing for instance a change of sign of the mixed charge-spin asymmetry and the mixed charge-spin difference for the MSTW case for instance. Note that the change is also huge in the case of GK parametrization but there is no sign change, indicating a significant model dependence. At any rate, NLO effects should be highly visible at COMPASS, which probes processes occuring at higher energy than JLab.

\subsection{Timelike Compton scattering}

\begin{figure}[ht]
\begin{center}
  \includegraphics[width=7 cm]{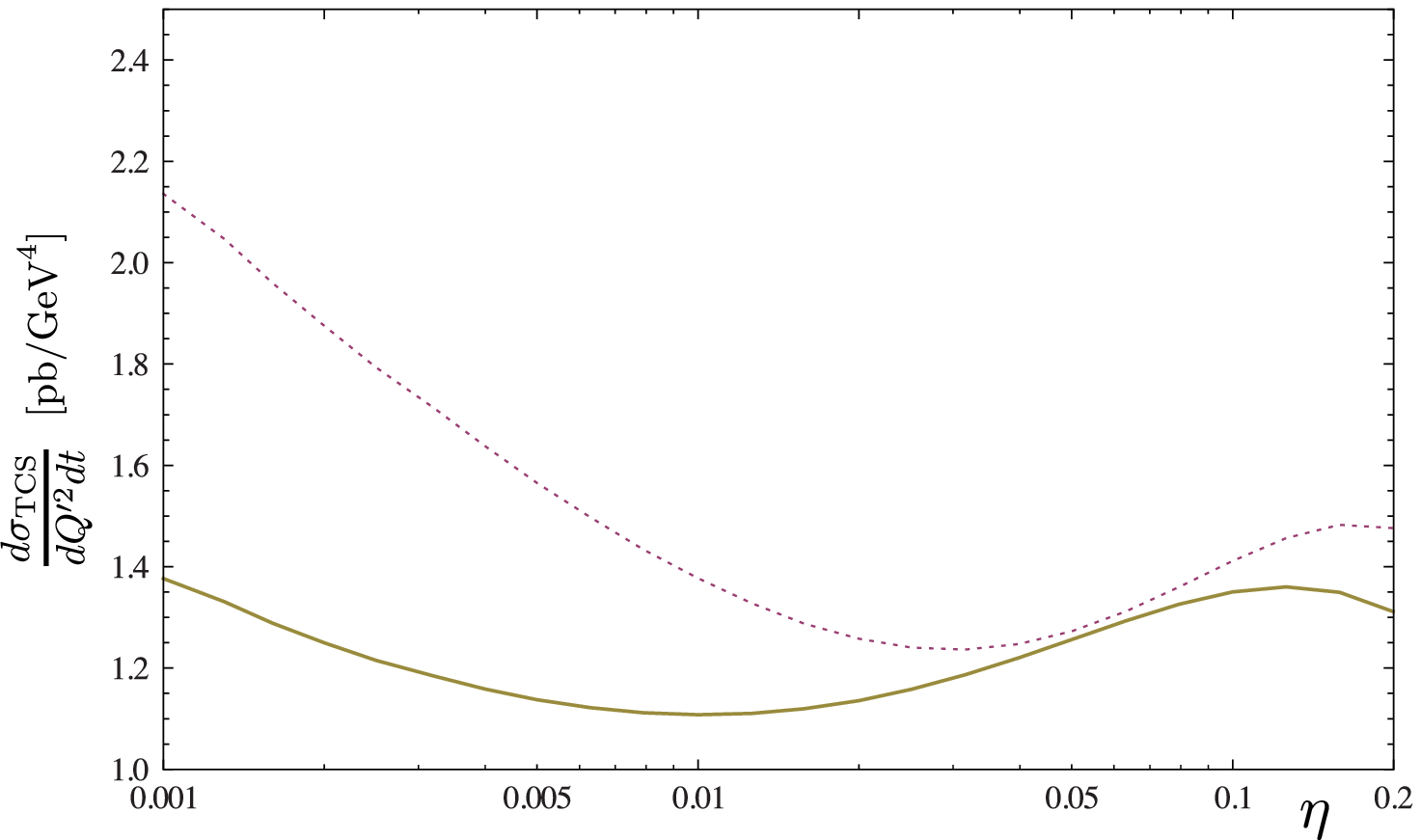}   ~~~~~~   
  \includegraphics[width=7 cm]{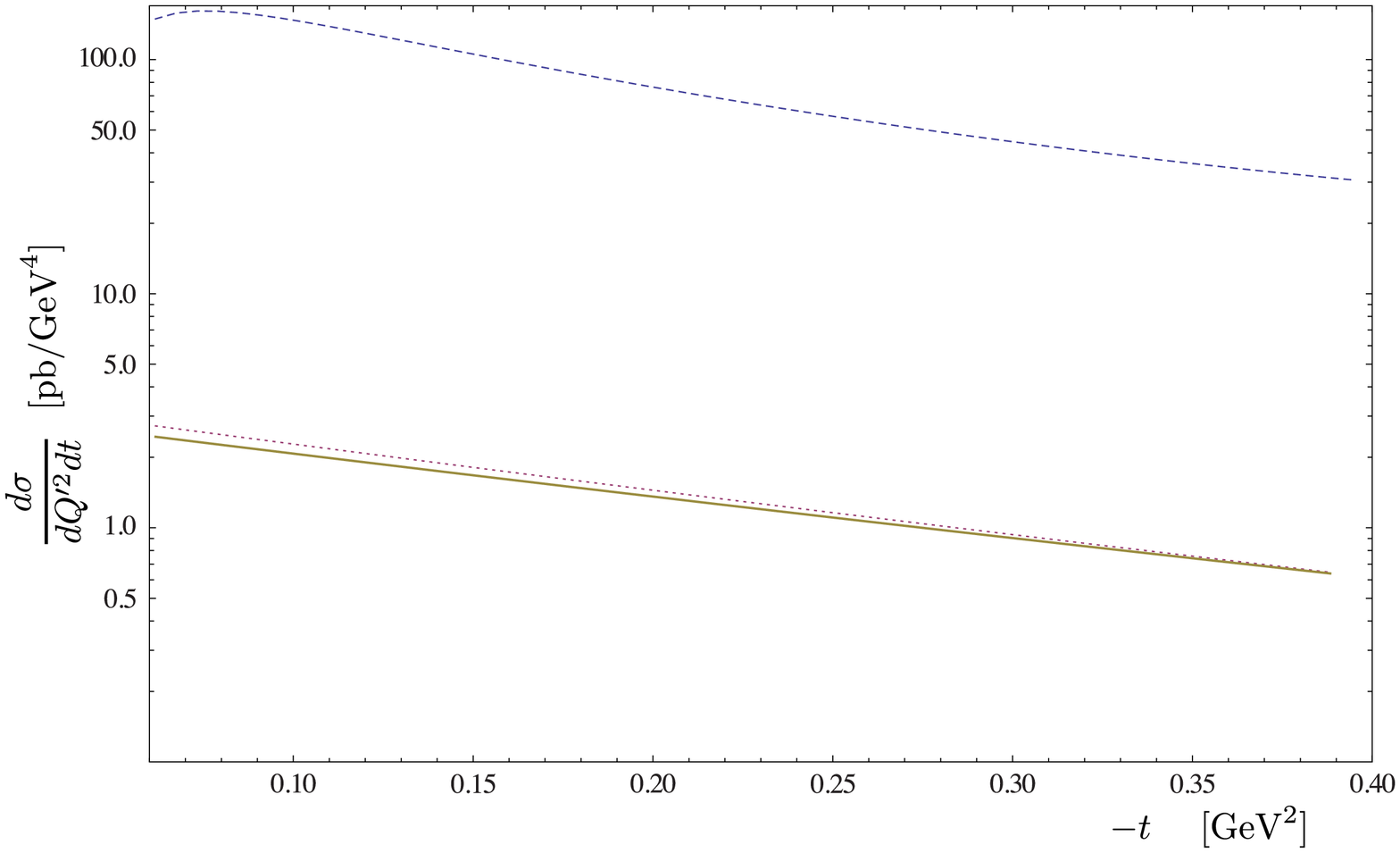}
\caption{left : LO (dotted) and NLO (solid)  the TCS contribution to the cross-section as a function of $\eta$ for  $Q^2 =  \mu^2 = 4$~GeV$^2$,  and $t= -0.2$~GeV$^2$ integrated over $\theta \in (\pi/4,3\pi/4)$ and over $\phi \in (0,2\pi)$. Right :  LO (dotted) and NLO (solid) TCS and Bethe-Heitler (dashed) contributions to the cross-section as a function of $t$ for  $Q^2 =  \mu^2$~GeV$^2$ integrated over $\theta \in (\pi/4,3\pi/4)$ and over $\phi \in (0,2\pi)$ for $E_\gamma=10$~GeV ($\eta \approx 0.11$).}
\label{fig:TCS_LO_NLO}
\end{center}
\end{figure}

\begin{figure}[ht]
\begin{center}
  \includegraphics[width=10 cm]{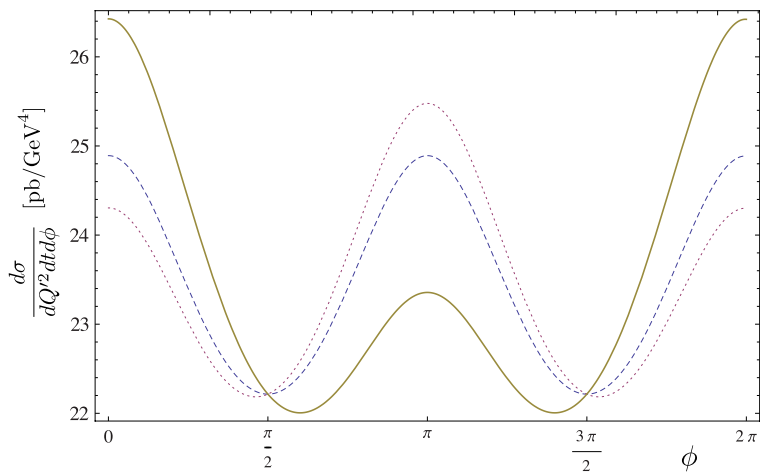}
\caption{The $\phi$ dependence of the cross-section at $E_\gamma = 10$ GeV,  $Q^2 =  \mu ^2 = 4$~GeV$^2$,  and $t= -0.1$~GeV$^2$ integrated over $\theta \in (\pi/4,3\pi/4)$: pure Bethe-Heitler contribution (dashed), Bethe-Heitler plus interference contribution at LO (dotted) and NLO (solid). }
\label{fig:xsec_phidep}
\end{center}
\end{figure}

Now we pass to predictions for the observables in the timelike counterpart of DVCS, namely TCS. On  the left part of Fig.~\ref{fig:TCS_LO_NLO} we show the TCS contribution to the differential cross section as a function of the skewness $\eta$ for  $Q^2 =  \mu^2 = 4$~GeV$^2$,  and $t= -0.2$~GeV$^2$ integrated over $\theta \in (\pi/4,3\pi/4)$ and over $\phi \in (0,2\pi)$. We see that the inclusion of the NLO corrections is more important at small skewness. On the right panel of Fig.~\ref{fig:TCS_LO_NLO} we show that the Bethe-Heitler dominates the integrated cross-section for this kinematics. In consequence, more differential observables, as the azimuthal $\phi$ dependence (with angles $\theta$ and $\phi$ defined in Ref.~\cite{BDP} ) reveal in a better way the different contributions. Moreover simple $\phi$ dependence of the interference term allows for an easy access to the real part of the CFFs, which as we observed on the Fig.~\ref{fig:TCSRe2x2}, are subject to the big NLO corrections. We indeed observe that effect on the Fig.~\ref{fig:xsec_phidep}, which shows the $\phi$ dependence of the unpolarized differential cross sections for pure BH process, and with a LO and NLO corrections to the interference term.

\begin{figure}[ht]
\begin{center}
  \includegraphics[width=10 cm]{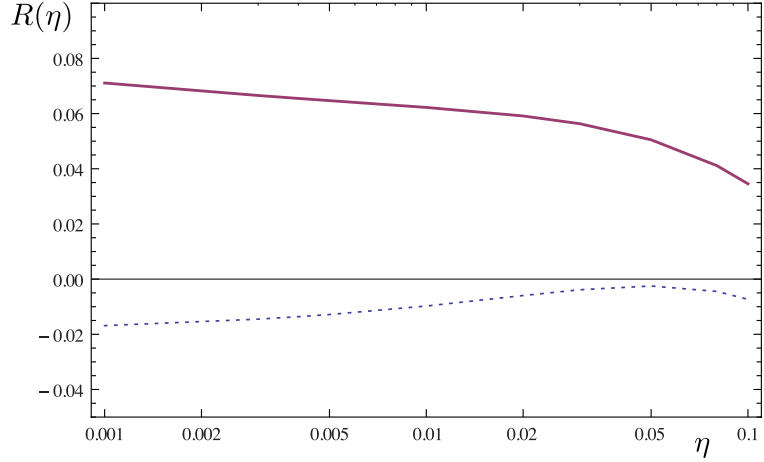}
\caption{Ratio R defined by Eq.~(\ref{eq:R_ratio}) as a function of $\eta$, for $Q^2 = \mu_F^2 = 4$~GeV$^2$ and $t= -0.1$~GeV$^2$. The dotted line represents LO contribution and the solid line represents NLO result.  }
\label{fig:Ratio_R}
\end{center}
\end{figure}


To quantify how big is the deviation from pure Bethe-Heitler process in the unpolarized cross section we calculate (see Fig.~\ref{fig:Ratio_R}) the ratio R, defined in Ref.~\cite{BDP} by
\begin{equation}
R(\eta) =  \frac{2\int_0^{2 \pi} d \varphi \,\cos \varphi\, \frac{dS}{dQ'^2dtd\varphi}}{\int_0^{2 \pi} d \varphi\frac{dS}{dQ'^2dtd\varphi}}
\,,
\label{eq:R_ratio}
\end{equation}
where $S$ is a weighted cross section given by Eq.~(43) of Ref.~\cite{BDP}. It is plotted on Fig.~\ref{fig:Ratio_R} as a function of the skewness $\eta$  for  $Q^2 =  \mu^2 = 4$~GeV$^2$,  and $t= -0.2$~GeV$^2$. In the leading twist the numerator is linear in the real part of the CFFs, and the denominator, for the kinematics we consider, is dominated by the Bethe - Heitler contribution. The inclusion of NLO corrections to the TCS amplitude is indeed dramatic for such an observable and includes also change of sign. 



\begin{figure}[ht]
\begin{center}
  \includegraphics[width=0.4\textwidth]{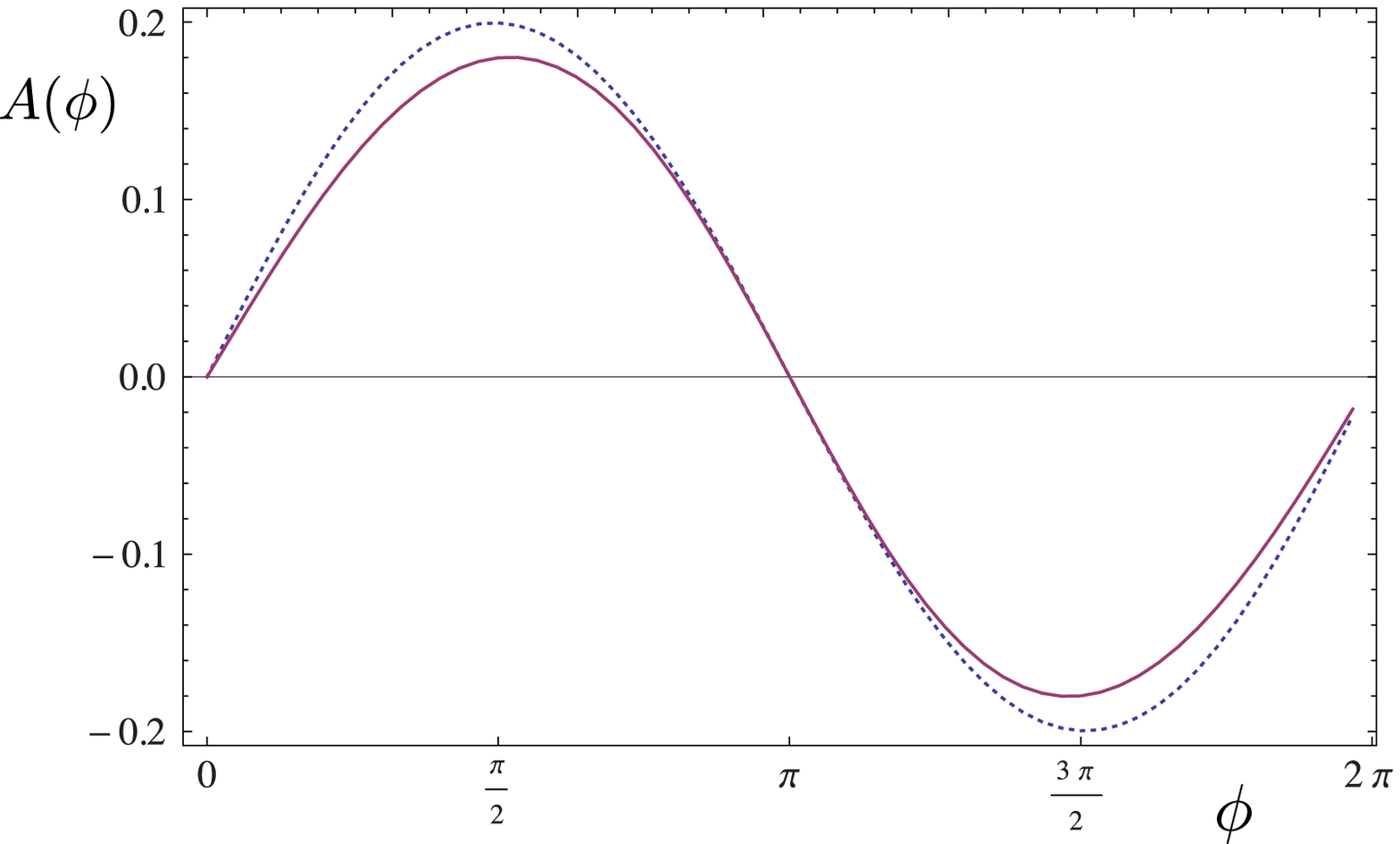}
  \hspace{0.05\textwidth}
  \includegraphics[width=0.4\textwidth]{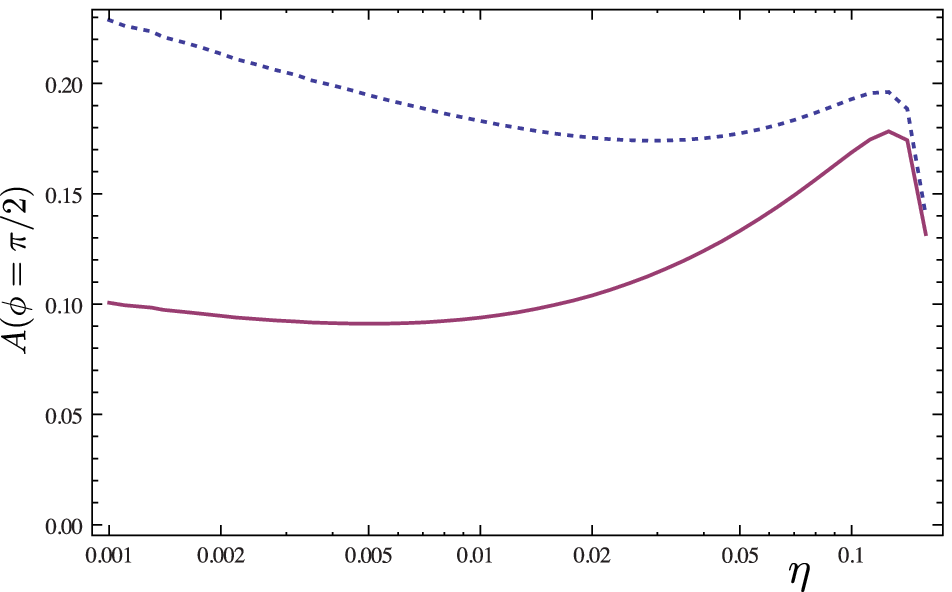}
\caption{
(Left) Photon beam  circular polarization asymmetry as a function of $\phi$, for $t=-0.1$~GeV$^2$, $Q^2 =  \mu^2 = 4$~GeV$^2$, integrated over $\theta \in (\pi/4,3\pi/4)$ and for $E_\gamma = 10$~GeV ($\eta \approx 0.11$).
(Right) The $\eta$ dependence of  the photon beam  circular polarization asymmetry for  $Q^2 =  \mu^2 = 4$~GeV$^2$,  and $t= -0.2$~GeV$^2$ integrated over $\theta \in (\pi/4,3\pi/4)$. The  LO result is shown as the dotted line, the full NLO result by the solid line.}
\label{fig:Asymmetry_xi}
\end{center}
\end{figure}

Imaginary parts of the CFFs are accesible through observables making use of photon circular polarizations \cite{BDP}. The photon beam circular polarization asymmetry
\begin{equation}
A= \frac{\sigma^+ - \sigma^-}{\sigma^+ + \sigma^-}\,,
\end{equation}
is shown on the left part of Fig.~\ref{fig:Asymmetry_xi}, as a function of $\phi$ for the kinematic variables relevant for JLab: $Q^2 =4$~GeV$^2$= $\mu_F^2$, $t=-0.1$~GeV$^2$ and $E_\gamma = 10$~GeV (which corresponds to $\eta \approx 0.11$). The same quantity is shown on the right panel of Fig.~\ref{fig:Asymmetry_xi} as a function of $\eta$ for $\phi = \pi/2$ and $Q^2 =4$~GeV$^2$= $\mu_F^2$. The effect of the NLO corrections on that observable is rather large, ranging from $10\%$ at the $\eta=0.1$ (relevant for JLab) through $30\%$ at $\eta=0.05$ (relevant for COMPASS) up to $100\%$ at very small $\eta$'s.

\section{Conclusion.}
Deeply Virtual Compton Scattering, both in its spacelike and timelike realizations, is the golden channel to extract GPDs from experimental observables. This extraction may be seen as a two step process: firstly, Compton form factors may be separated from a careful analysis of various differential cross sections and asymmetries. Secondly, convolutions of coefficient functions with model GPDs may be confronted to these CFFs. We have demonstrated here, in the case of medium energy kinematics which will be explored in the near future at JLab and COMPASS,  that the inclusion of NLO corrections to the coefficient function was an important issue, and that the difference of these corrections between the spacelike and timelike regimes was so sizeable that they can be promoted to the status of direct tests of the QCD understanding of the reactions.

Let us stress again a feature that was largely overlooked in previous studies, namely the importance of gluon contributions to the DVCS amplitude, even when the skewness variable $\xi$ is in the so-called valence region. This is not a real surprise when one recalls that gluons (in terms of distribution functions) are by no means restricted to the very low $x$ region and that gluon CFFs at a given $\xi$ value also depend on gluon PDFs at lower values of $x$. This effect is particularly big when one considers the real part of CFFs in the timelike case. This promotes the observables related to this quantity as sensitive probes of the 3-dimensional gluon content of the nucleon.

We did not extend our study to the very high energy regime which will be explored (both for spacelike and timelike Compton scattering) at future electron-ion colliders, nor to the case of ultraperipheral collisions at present hadron colliders. This will be addressed in separate studies. We did not discuss the rich scope of factorization scale dependence issues, which deserves special attention. Contrarily to the case of inclusive reactions where various strategies have been built to optimize the factorization scale, it has been shown in Ref.~\cite{Anikin} that it was quite impossible to find a recipe to minimize higher order corrections both for the real and the imaginary part of a CFF. Moreover we find it difficult to advocate a different choice of scale for the two cases of timelike and spacelike reactions.

\section*{Acknowledgements} 
We are grateful to Markus Diehl, Dieter M\"{u}ller,  Stepan Stepanyan, Pawe{\l} Nadel-Turo\'nski and Samuel Wallon for useful discussions and correspondence. This work is partly supported by the Polish Grants NCN No DEC-2011/01/D/ST2/02069, by the Joint Research Activity "Study of Strongly Interacting Matter" (acronym HadronPhysics3, Grant Agreement n.283286) under the Seventh Framework Programme of the European Community, by the GDR 3034 PH-QCD, and the ANR-12-MONU-0008-01, and by the COPIN-IN2P3 Agreement.

\end{document}